\documentclass[12pt]{article}
\usepackage[pdftex]{graphicx}

\begin{document}

\title{Bragg Reflection Waveguide: Anti-Mirror Reflection and Light Slowdown}
\maketitle

G. G. Kozlov, V. S. Zapasskii, Yu. V. Kapitonov, and V. V. Ovsyankin

Institute of Physics, Saint-Petersburg State University,
Saint-Petersburg, 198504, Russia
\vskip10mm
\begin{abstract}
The effect of the light group velocity reduction in dielectric Bragg reflection waveguide structures
(SiO$_2$/TiO$_2$) in the vicinity of the cutoff frequency is studied experimentally. The effect of
anti-mirror reflection, specific for the Bragg reflection waveguides, is described and employed for
detection of "slow light". The experiments were performed with the use of the Ti:sapphire laser pulses
~ 100 fs in length. The group index $n_g \sim$ 30 with a fractional pulse delay (normalized to the
pulse width) of $\sim$ 10 is demonstrated. The problems and prospects of implementation of the
slow-light devices based on the Bragg reflection waveguide structures are discussed.
\end{abstract}

\section{INTRODUCTION}

Engineering of optical materials with anomalously high group velocity index ($n_g >> 1$) constitutes the
main content of the field of optics that arose in the last decade and is referred to as "slow light"\cite{Sl}.
Importance of the research in the framework of this trend is related to practical demand for the
compact time delay lines for the information processing systems. Possibility of such applications
imposes certain requirements to the slow-light systems, the most important among them being large
bandwidth ($\sim 10^9$ Hz) and room-temperature operation. From the viewpoint of these requirements, it
seems highly promising to create the slow-light devices using photonic crystals and related waveguide
structures \cite{Kra,Tos}.
 In recent years a number of papers have been published devoted to the simplest system of this type,
namely, to the classical Bragg reflection waveguide\cite{Koyama1,Koyama2}.
 A theoretical analysis of the Bragg reflection waveguides \cite{Koz1}, aimed at studying the
possibility of their use for obtaining slow light, has shown that, in principle, this system allows one
to achieve the group index $n_g \sim$ 1000 for the spectral width of the optical pulse $\sim 10^9$ Hz with a
fractional delay (i.e., the delay normalized to the pulse width) of $\sim$ 20. In this case, the
requirements to quality of the Bragg structure prove to be compatible with capabilities of the modern
technologies. The goal of this study is to demonstrate experimentally the optical pulse group delay in
a Bragg reflection waveguide. Though the quality of the waveguides used in this study was not high
enough, the experience accumulated in the course of the work with these structures and the effects we
observed allow us to point out a number of specific problems that arise when working with Bragg
reflection waveguides and to find the ways of their solution.

Remind briefly the main properties of the Bragg reflection waveguide structures \cite{BW,Koz1}. The Bragg
reflection waveguide is comprised of two parallel Bragg mirrors separated by a guiding core with a
thickness of 2$l$ filled with a dielectric with the refractive index $n_0$. Each Bragg mirror represents
a periodic structure of dielectric layers, with its period formed by a pair of layers with different
refractive indices ($n_2 > n_1$). Optical thicknesses $n_1l_1$ and $n_2l_2$ of these layers are chosen to
maximize reflectivity at a specified frequency of the electromagnetic wave $\bar\omega$ and are equal to a
quarter of the appropriate wavelength, i.e., $l_in_i = \pi c/2\bar\omega, i = 1, 2$. At $n_2/n_1 \sim 2$
or greater, reflectivity of the Bragg mirror rapidly increases with increasing number of periods $N$,
and, at $N \sim$ 6 - 8, may reach 99$\%$.

The light slowdown in the Bragg reflection waveguide may be considered as a result of its zigzag-like
propagation in the guiding core, when the electromagnetic wave experiences successive reflections from
the Bragg mirrors. Such a regime is characterized by the wave vector $q$ lying in the plane of the
waveguide layer. If the optical thickness of the guiding core is chosen equal to a half-wavelength
corresponding to greatest reflectivity of the Bragg mirrors ($ln_0 = \pi c/2\omega$), and the mirrors
are perfectly flat and thick ($N \rightarrow \infty$), then the wave vector $q$
 is real and is related to the wave
frequency $\omega$ by the following law of dispersion  \footnote { This formula is valid when
$\omega$ is close to $\bar\omega$. For the study of slow light, only this case is of interest.}
\begin{equation}
q=G\sqrt{\omega-\bar\omega}
\label{1}
\end{equation}

where the constant $G$ is determined by specific values of parameters of the structure \cite{Koz1}, and
$\bar\omega$
is the "cutoff frequency" of the waveguide: the waves with frequencies below $\bar\omega$ cannot propagate
in such a waveguide. Due to imperfections of the Bragg mirrors, the wave vector $q$ acquires an
imaginary increment, and the considered regime of propagation of the electromagnetic wave becomes
damping.

If a spectrally narrow optical pulse with the carrier frequency $\omega$ is introduced into the
waveguide, then, as is known, such a pulse (or, more exactly, its envelope) will propagate in the
waveguide with the group velocity $v_g$ governed by the relationship

\begin{equation}
v_g={c\over n_g}, \hskip5mm n_g=c{dq\over d\omega}={cG\over2 \sqrt{\omega-\bar\omega}}
\label{2}
\end{equation}

 where $n_g$ is the group refractive index. It is seen from this formula that as the carrier frequency
 $\omega$ approaches the cutoff frequency $\bar\omega$, the group index increases infinitely. This fact
 arouses interest to the structure under study from the viewpoint of the slow light.

The experiments described below were performed with the Bragg reflection waveguide structures formed by
the classical pair SiO$_2$/TiO$_2$, grown on a glass-ceramic substrate. The guiding core was made of SiO$_2$,
so that $n_0 = n_1 =$ 1.45 and $n_2 =$ 2.2 -- 2.3.

Our attempts to couple light into the Bragg reflection waveguide by focusing it onto the butt end of
the structure, which is common for the waveguides of total internal reflection, have failed. This can
be understood in terms of a simple geometrical model. The angle of incidence $\phi$ onto the Bragg
mirrors (Fig. 1) at the frequency of the wave $\omega$ close to the cutoff frequency $\bar\omega$, for
operating mode of the electromagnetic wave, appears to be small (at $\omega = \bar\omega$, it turns into
zero). On the contrary, the angle of incidence of the operating mode onto the plane of the butt end
(equal to 90$^o - \phi$, see Fig. 1) appears to be large and may become greater than the angle of
total internal reflection, which prevents the light from coming out of the waveguide. For the same
reason, injection of the light through the butt end is hampered. For instance, if the light frequency
$\omega$ exceeds by 5$\%$ the cutoff frequency $\bar\omega$, then the angle of incidence of the operating mode
upon the Bragg mirrors $\phi$ in the waveguide appears to be equal to 30$^o$, which corresponds to the
angle of incidence of the operating mode onto the plane of the butt end of the waveguide $\sim$ 60$^o$, which,
in turn, exceeds the angle of total internal reflection even for the low-index component of the Bragg
structure (43$^o$ for SiO$_2$).

In virtue of the aforesaid, the experiments on registration of the group delay in the Bragg reflection
waveguides were performed on the samples with locally thin (see below) Bragg mirror, which provided
possibility of local coupling of light into the waveguide. In this case, we employed the phenomenon of
anti-mirror reflection, specific for the waveguiding structures with leakage described in the next
section.

\section{ANTI-MIRROR REFLECTION}

Consider the Bragg reflection waveguide with the mirrors whose reflectivities $r$ are sufficiently high.
If a plane monochromatic wave ($A\exp[\imath {\bf kr}], |{\bf k}|\equiv k = \omega/c$, $c$
 is the velocity of light in vacuum)
with the frequency $\omega > \bar\omega$ is incident at an angle of $\alpha$ from the upper semispace
(vacuum), then the field in each layer of the structure can be represented as a sum of two plane waves
with the same horizontal components of the wave vector $q_x = k\sin\alpha \equiv q$ and with equal in
magnitude, but opposite in sign vertical components $q_z=\pm\sqrt{k^2n^2-q^2}$
($n$ is the refractive index of the
layer under study). In the general case, the strength of the field rapidly decreases with the depth of
the structure. The transmission $|T/A|^2$, in this case, proves to be small, while the reflection
$|R/A|^2$, on the contrary, approaches unity. However, if the angle of incidence $\alpha$ and the
frequency $\omega$ are chosen so that condition (\ref{1}) is satisfied
\begin{equation}
q={\omega\over c}\sin\alpha=G\sqrt{\omega-\bar\omega},
\label{3}
\end{equation}
then the electromagnetic field strength in the guiding core resonantly increases, and, in spite of the
fact that this core is separated from the upper and lower semispaces by the Bragg mirrors with high
reflectivities, the field in it starts to play decisive role in formation of the reflected and
transmitted waves. Indeed, when the resonance condition (\ref{3}) is satisfied, there arise two waves in the
guiding core $\exp[\imath(qx\pm\sqrt{k^2n_0^2-q^2} z)]$ with great amplitudes
большой ($\sim A(1-r)^{-1/2}$). \footnote {The regime corresponding
to the two indicated waves was called above zigzag-like propagation.} Since the Bragg mirrors
surrounding the guiding core have a finite number of periods, propagation of this waves is accompanied
by the leakage, which results in appearance of the waves, in the upper and lower semispaces, that
travel outward from the Bragg structure. Due to a large value of the field in the guiding core,
amplitude of these waves appears to be close to $A$. As a result, when the resonance condition (\ref{3}) is
satisfied, the transmissivity $|T/A|^2$ proves to be close to unity. Amplitude of the reflected wave $R$,
in this case, is determined by the interference of the wave arising due to leakage from the guiding
core and the wave arising due to reflection of the incident wave from the upper Bragg mirror. In
conformity with the energy conservation law, this interference is of destructive nature, which is
revealed in resonant decrease of the reflected wave amplitude $R$.

The above treatment is valid for the Bragg structures {\it infinite} along the $x$-axis. Assume now that we
deal with a {\it semi-infinite} Bragg reflection waveguide, with a plane monochromatic wave satisfying
resonance condition (\ref{3}) incident upon it. As in the case of infinite waveguide, in the guiding core
there will arise the waves of large amplitude  ($\sim A(1-r)^{-1/2}$) with the $x$-component of the wave
vector equal to that for the incident wave: $q = (\omega/c)\sin\alpha$. Specific property of the
semi-infinite waveguide is that these waves, propagating in the guiding core from the left to the
right, should necessarily be reflected at the right butt end of the waveguide (Fig. 1). As a result, in
the guiding core, there will arise the waves traveling in the opposite direction and having the
$x$-component of the wave vector equal to $-q$. The leakage of these waves will result in appearance, in
the lower and upper semispaces, of the waves with the $x$-component of the wave vector equal to $-q$,
which are denoted in Fig. 1 as $R'$ and $T'$, respectively. In the upper semispace, the wave $R'$ (Fig. 1)
travels exactly towards the incident wave. In the lower semispace, this effect leads to appearance of
the refracted wave $T'$ propagating in the direction antiparallel to the direction of the reflected wave
$R$. Thus, the resonant scattering of the plane wave by the semi-infinite Bragg reflection waveguide is
accompanied by appearance of the {\it anti-mirror} wave $R'$ in the upper semispace and the refracted wave
$T'$ in the lower semispace. Note that a similar effect of anti-mirror reflection in plane crystalline
systems with excitonic type of excitation was described in \cite{Koz2}.

It is noteworthy that the waves with $x$-component of the wave vector equal to $q$ are undamped: the
energy loss due to the leakage (forming normal reflected and transmitted waves) is compensated by the
energy income from the incident wave through the upper Bragg mirror. The waves with the $x$-component of
the wave vector equal to $-q$ do not have such a channel for the loss compensation. For this reason,
the leakage of these waves leads to a decay of their amplitude when moving outward from the right end
of the waveguide (Fig. 1). This decay can be characterized by the length $L$, defined as a distance from
the right end of the waveguide to the point where the amplitude of the reflected wave decreases by a
factor of $e$. \footnote { The decay length can be expressed as the inverse imaginary part of the
complex wave vector entering the dispersion law of the real Bragg reflection waveguide \cite{Koz1}.} The decay
length is determined by reflectivity of the Bragg mirrors $r$ and increases with $r$ tending to unity.
Thus, the anti-mirror reflection described above can be formed only by a stripe of $\sim L$ in width near
the edge of the waveguide and, therefore, is characterized by a finite divergence of the order of
$\sim\lambda/L$ ($\lambda$ is the appropriate light wavelength).

To observe the effect of anti-mirror reflection, we used the Bragg reflection waveguide with its mirror
comprised of 7 pairs of quarter-wave layers SiO$_2$/TiO$_2$. The cutoff frequency $\bar\omega$ corresponded to
the wavelength $\lambda = 2c/{\bar\omega} = $820 nm. Schematic of the experimental setup is shown in Fig.
2. The Bragg reflection waveguide 1 mounted on a special holder executed rotary vibrations around the
axis passing through the edge of the waveguide. The laser beam reflected by beamsplitter 2 was
directed to the sample, which could be inserted into the beam using micropositioner 5. A photodetector
(PMT) 3 with attenuating aperture 4 was placed at the opposite side of the beamsplitter. The aperture
was placed along the line that continued the beam incident on the Bragg reflection waveguide (Fig. 2).
Photocurrent of the PMT was detected by oscilloscope 6 whose scan was synchronized with the swinging
motion of the waveguide. This setup allowed us to detect scattering in the anti-mirror direction as a
function of the angle of incidence of the laser beam onto the Bragg waveguide. The experiments have
shown that, at the angle of incidence determined by Eq. (\ref{3}), this dependence exhibited a sharp peak,
which could be observed only when the laser beam hit the edge of the waveguide. The anti-mirror
reflection, in our experiments, proved to be highly directional which was indicated by a sharp drop of
the detected photocurrent upon displacement of the input aperture of the photodetector with respect to
the anti-specular direction (Fig. 2). The appropriate angular width is estimated to be $\sim$ 0.01 rad
\footnote { This value corresponds to divergence in the plane of incidence; in the orthogonal
direction, the divergence was by an order of magnitude larger, which we ascribe to unevenness of the
waveguide edge.}

From all the aforesaid, one can conclude that the effect of anti-mirror reflection is related to
{\it propagation} and {\it reflection} of the electromagnetic waves localized in the guiding core of the Bragg
reflection waveguide, which allows one to use this effect for studying the light pulse slowdown in
these structures.

\section{OBSERVATION OF SLOW LIGHT IN THE BRAGG REFLECTION WAVEGUIDE}

In the experiments on observation of the light pulse group delay, we used a sample of the Bragg
reflection waveguide with mirrors consisted of 18 pairs of the quarter-wave layers SiO$_2$/TiO$_2$. For
this sample, we have measured experimentally dependence of the resonance frequency $\omega$ on the
angle of incidence $\alpha$, have found the cutoff frequency and the constant $G$ entering Eqs. (\ref{1}) --
(\ref{3}). We have also plotted dependences of the group refractive index $n_g$ on the wavelength
$\lambda$ (Fig. 3).

The sample was prepared as follows. In the upper Bragg mirror of the waveguide, not far from its edge,
we etched a pit of 3 $\times$ 1 mm$^2$ in size and $\sim$ 2 $\mu$m in depth, \footnote { The etching was
performed during 10 -- 15 s in 37$\%$ hydrofluoric acid diluted by water in the volumetric ratio 1:5.}
which corresponded to local reduction of thickness of the upper mirror approximately by a factor of 2
(Fig. 4a). Then, the edge of the waveguide, nearest to the pit, was grounded so that the distance
between the edge and the nearest slope of the pit (the length of segment $|BC|$ in Fig. 4a) was $\sim$
 60 -- 70 $\mu$m.

To observe the group delay, a short laser pulse ($\sim$ 100 fs) was focused by a long-distance lens ($f
=$ 150 mm) onto the slope of the pit nearest to the edge of the sample in the upper Bragg mirror (point
$A$ in Fig. 4a). The sample was oriented so that the angle of incidence $\alpha$ and the carrier frequency
of the pulse $\omega$ satisfied the resonance condition (\ref{3}). Through the locally thin Bragg mirror, the
pulse penetrated into the guiding core, traveled along it to the right edge of the structure, reflected
from it, and came out of the structure in the area of the pit, forming the pulse of anti-mirror
reflection described in the previous section. In this case, contribution into the anti-mirror
reflection connected with the leakage of the light propagating from the edge of the sample to the slope
of the pit (segment $|BC|$ in Fig. 4a) appeared to be strongly suppressed due to high reflectivity of the
waveguide mirrors in this region.
 Along with the pulse of anti-mirror reflection, we always observed, in the experiment, a pulse
reflected directly from the edge of the pit (point $A$ in Fig. 4a), which we used as a zero-delay pulse.
By measuring the delay $\Delta T$ between the zero-delay pulse and the pulse of anti-mirror reflection,
we could estimate the group velocity as $v_g = 2|BC|/\Delta T$. These measurements were performed with
the aid of the Michelson interferometer using the following method.

 Schematic of the setup is shown in Fig. 5. The input beam of the mode-locked Ti-sapphire laser was
split by beamsplitter 3 into two beams. One of them (signal beam) was directed to sample 1 through
the long-distance lens 2. The second beam (reference) was directed to mirror 11 mounted on
micropositioner 12. Photocurrent of the PMT 5 placed at the exit of the Michelson interferometer was
governed by interference of the reference and signal beams reflected, respectively, by mirror 11 and
sample 1. To increase the contrast of the interference, the aperture 4 with a diameter of $\sim$ 50 -- 100
$\mu$m was placed in front of the PMT. Since the input laser beam represented a succession of short
(spatially localized) light pulses, the interference could be observed only when the time interval
between the signal and reference pulses (which, in the general case, passed different path lengths in
different arms of the interferometer) was so small that pulses exhibited temporal (and spatial)
overlap.
The optical path length in the reference channel was subjected to a weak sine modulation at the
frequency $\nu$ using vibrator 10. As a result, the frequency of the interference signal on the PMT,
related to scanning of the path difference in the arms of the interferometer was modulated at the
frequency $2\nu$. Using a differentiating circuit (Fig. 5), the frequency-modulated signal was
transformed into the amplitude-modulated, which, after amplification (6) and quadratic detection was
transformed into ac voltage with the frequency $2\nu$. This voltage, proportional to the signal of
interference of the reference and signal pulses, was detected using the lock-in amplifier 7, whose
reference signal (with the frequency $2\nu$) was obtained by full-wave rectification of output voltage
of the audio-frequency oscillator 8 controlling vibrator 10.

The measurements were performed as follows. As was already said, the optical pulse incident upon the
Bragg reflection waveguide sample (Fig. 4) gave birth to two reflected pulses: the zero-delay pulse
reflected from the slope of the pit (point $A$ in Fig. 4) and the group-delayed pulse of anti-mirror
reflection. For this reason, the initial position of mirror 11 in the reference arm of the
interferometer was chosen so that the delay of the reference pulse was slightly smaller than the delay
of the aero-delay pulse in the signal arm. After that, micropositioner 12 with mirror 11 was
uniformly translated by a stepping motor in the direction corresponding to increasing delay of the
reference pulse (to the right, in Fig. 5). The interference signal (output of the lock-in amplifier)
was recorded as a function of displacement of mirror 11. When the delay in the reference arm increased
so that the reference and zero-delay pulses overlapped, the first signal appeared at the output of the
lock-in amplifier 7. As the delay further increased, the interference signal disappeared and arose
again only when the reference pulse and the pulse of anti-mirror reflection overlapped. Thus, the
dependence of the interference signal on position of mirror 11 contained two peaks corresponding to
the zero-delay pulse and to the pulse of anti-mirror reflection from the Bragg structure. By measuring
displacement $l$ of mirror 11, corresponding to these two peaks, we could find group index of the
structure under study: $n_g = l/|BC|$, where $|BC|$ is the distance from the slope of the pit in the
waveguide to its edge (Fig. 4a).

The results of the measurements are presented in Fig. 6, which shows experimental dependences of the
interference signal on displacement of mirror 11 (Fig. 5) for three different carrier frequencies of
the probe pulses (with the wavelengths 765 nm (a), 780 nm (b) and 790 nm (c)). The left narrow peak in
all plots corresponds to zero-delay pulse. The presence of two broad peaks in Fig. 6b is related, as we
believe, to a stepwise unevenness of the pit edge in the region of focal spot of the incident beam (see
Fig. 4b). Because of this unevenness, the value $|BC|$ for different parts of the beam could be
noticeably different. Appearance of two or more peaks in the anti-mirror reflection was not rare in our
experiments. This is why we considered necessary to present at least one plot of this type (Fig. 6b).
After realignment of the scheme, it was usually possible to obtain the plot with a single anti-mirror
peak (Fig. 6d). As seen from Fig. 6, the fractional pulse delay is here $\sim$ 10.
For all the above dependences (Fig. 6), we calculated the group refractive indices. The results thus
obtained are shown by solid circles in Fig. 3, where we also present the expected spectral dependence
of the group refractive index for the structure under study. For $\lambda$ = 780 nm (Fig. 6b), we show
the results corresponding to both peaks. In all cases, the length $|BC|$ was taken to be 60 $\mu$m. This
estimate was made using the microphotograph of the appropriate region of the waveguide structure
presented in Fig. 4b. Focal spot of the incident light was positioned in the lower right corner of the
pit (Fig. 4b) and was ~ 150 $\mu$m in size. Taking into account low quality of the pit edge (Fig. 4b),
the agreement of the measured and calculated values of the group refractive indices can be considered
as good.

\section{DISCUSSION AND CONCLUDING REMARKS}

In spite of the fact that the value of the group index ($n_g$ = 30) achieved in this paper was not high
enough, the above experiments demonstrate good prospects of using the Bragg reflection waveguide
structures in the geometry of anti-mirror reflection for engineering slow-light devices and allow us to
formulate two problems that should be solved for successful development of this approach. The first of
them is related to the needed improvement of spatial uniformity of the waveguide structures. The second
one implies development of efficient technique for coupling light into the Bragg reflection waveguide.
Let us dwell upon these problems in more detail.

\subsection{Spatial nonuniformity of the Bragg reflection waveguide}

The distance 2$|BC|$ (Fig. 4a) passed by the optical pulses in our experiments was $\sim$ 120 $\mu$m. The
attempts to enlarge this distance to obtain greater absolute delays resulted in a sharp drop of the
anti-mirror reflection signal, which disagrees with our estimates of the pulse decay length for the
waveguides with mirrors consisting of $N$ = 18 pairs of the quarter-wave layers \cite{Koz1}. We believe that the
reason for the anomalously fast decay of the signal is related to the light scattering caused by
nonuniformity of the waveguide layer. As a result, the reflectivity of the Bragg mirrors appears to be
much lower than $r$ = 0.999999, corresponding to the ideal mirror with $N$ = 18. This is confirmed by the
relatively low Q-values of the transmission and reflection resonances ($\omega/\Delta\omega$ ), observed
in our samples, which should be  $\sim (1-r)^{-1}\sim 10^{6} - 10^{7}$.
 The large number of periods in the mirrors
($N$ = 18) resulted only in suppression of the amplitudes of the resonance peaks in the reflection and
transmission, but did not reduce their spectral widths, which, in our samples, corresponded to Bragg
mirrors with $N$ = 7 -- 8. A considerable nonuniformity of our structures also well correlated with
noticeable spatial fluctuations of the cutoff frequency.
As far as we know, the up-to-date technologies make it possible to obtain mirrors with the reflectivity
(in specular component) $\sim$ 0.99999. So high reflectivity can be achieved only with extremely uniform
mirrors, because nonuniformity of the structure inevitably leads to scattering of a fraction of the
light. We have every reason to suppose that the Bragg waveguide with such mirrors will make it possible
to observe propagation of optical pulses over the distances of the order of several mm \cite{Koz1}.

\subsection{The problem of input/output coupling}

As was already mentioned, the simplest and spectrally nonselective method for coupling light into the
waveguide by focusing it onto the butt end of the waveguide proves to be unacceptable. The method of
light coupling through the locally thin Bragg mirror used in the present study is spectrally selective,
since the frequency and the angle of incidence of the spectral component introduced into the waveguide
should be connected by relationship (\ref{3}). In this case, the spectral width $\Delta\omega$ of the
optical pulse that can be coupled into the waveguide is determined by finesse of the waveguide with the
locally thin mirror: $\omega/\Delta\omega\sim (1-r')^{-1}$, where $r$ is the reflectivity of the locally
thin Bragg mirror. In our experiments, $\omega/\Delta\omega \sim$ 1000, which caused a noticeable narrowing of
the backscattered pulse spectrum with appropriate increase in its spatial length as compared with the
spectrally undistorted zero-delay pulse. It is possible that further decrease in the local thickness of
the upper mirror (segment $|AB|$ in Fig. 4a) and a sharper focusing will allow us to reduce the spectral
distortions and intensity loss upon injection of the optical pulse into the waveguide. This possibility
should be studied using the Bragg structures with substantially higher quality of the pit slopes,
which, in our samples, were far from perfect (Fig. 4b).
We believe that, for injection of a broadband pulse into the Bragg waveguide, it will be efficient to
use a specially designed diffraction grating (placed in front of the waveguide or deposited upon it)
with such an angular dispersion that the frequency and angle of incidence of each spectral component of
the input beam, at least in linear approximation, will meet Eq. (\ref{3}). This technique could substantially
reduce spectral selectivity of the I/O system of the Bragg waveguide.

In conclusion, we would like to thank associates of the Vacuum sector of the Optical shop of OKB FIAN
(Troitsk) L. V. Perebyakina and L.P.Khraponov for preparation of samples of the Bragg reflection
structures.

\subsection{ACKNOWLEDGEMENTS}

The work was supported by the Analytical Departmental Targeted Program "Development of Scientific
Potential of Higher School" of 2009-2010 (project no. 2.1.1/1792).



\newpage
\begin{figure}
\begin{center}
\includegraphics [width=10cm]{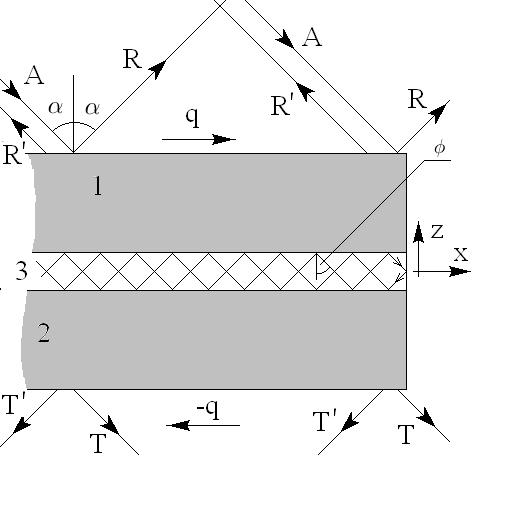}
 \caption{The origin of the anti-mirror reflection in a semi-infinite Bragg reflection waveguide. 1 and
2 -- upper and lower Bragg mirrors; 3 guiding core, the zigzag
shows the wave with the $x$-component equal to $q$ and the wave
arising due to reflection from the right edge of the waveguide
with the $x$-projection equal to $-q$; $A$ - incident wave; $R$
and $T$ --
 normal reflected and transmitted waves; $R'$
and $T'$ -- anti-mirror and refracted waves arising due to the
wave, in the guiding core, reflected from the butt end of the
waveguide and having the horizontal component of the wave vector
equal to $-q$.}
 \label{fig1}
 \end{center}
\end{figure}

\begin{figure}
\begin{center}
\includegraphics [width=10cm]{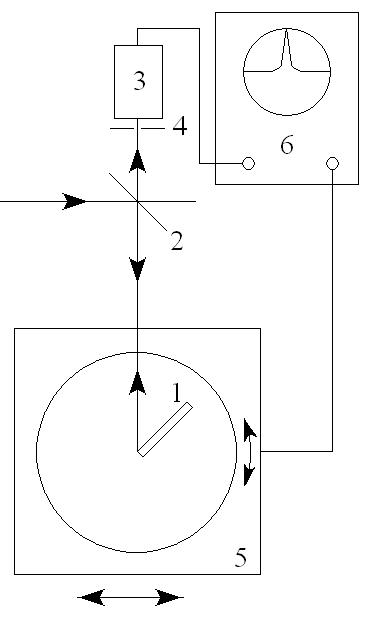}
 \caption{Schematic of the setup for observation of the anti-mirror reflection effect. 1 - Bragg
reflection waveguide, 2 - beamsplitter, 3 - PMT, 4 - aperture, 5 -
micropositioner, and 6 - oscilloscope. }
 \label{fig2}
 \end{center}
\end{figure}

\begin{figure}
\begin{center}
\includegraphics [width=10cm]{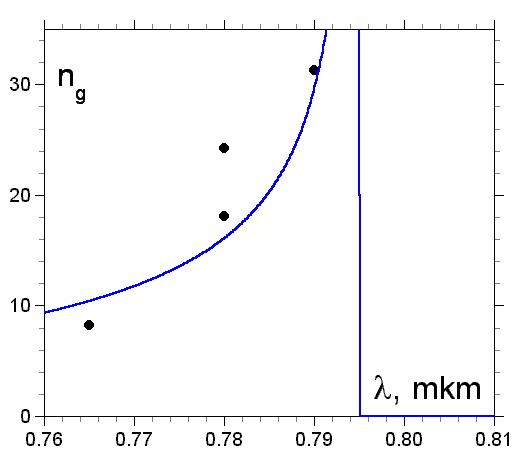}
 \caption{Solid curve - calculated spectral dependence of the group refractive index $n_g$ for the Bragg
waveguide under study, solid circles -- experimental values of
$n_g$ for the wavelengths 765 nm, 780 nm, and 790 nm. }
 \label{fig3}
 \end{center}
\end{figure}

\begin{figure}
\begin{center}
\includegraphics [width=10cm]{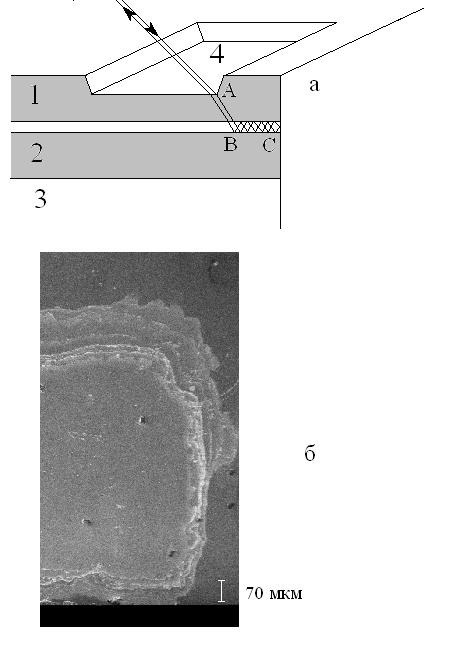}
 \caption{(a) - coupling light into the Bragg reflection waveguide through the locally thin upper mirror.
1 and 2 - upper and lower Bragg reflection mirrors, 3 - substrate,
4 - locally thin region. (b) - microphotograph of the locally thin
region of the Bragg reflection waveguide.  }
 \label{fig4}
 \end{center}
\end{figure}

\begin{figure}
\begin{center}
\includegraphics [width=10cm]{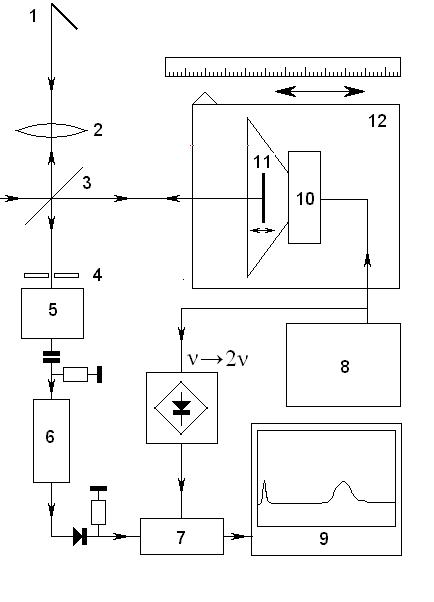}
 \caption{Schematic of the setup for observation of the group delay in the Bragg reflection waveguide in
the configuration of anti-mirror reflection. 1 - Bragg reflection
waveguide, 2 - long-distance lens, 3 - beamsplitter, 4 - aperture,
5 - PMT, 6 - amplifier, 7 - lock-in amplifier, 8 -
audio-oscillator, 9 - PC, 10 - vibrator (loudspeaker), 11 --
mirror of the reference arm of the interferometer, 12 -
micropositioner. }
 \label{fig5}
 \end{center}
\end{figure}

\begin{figure}
\begin{center}
\includegraphics [width=10cm]{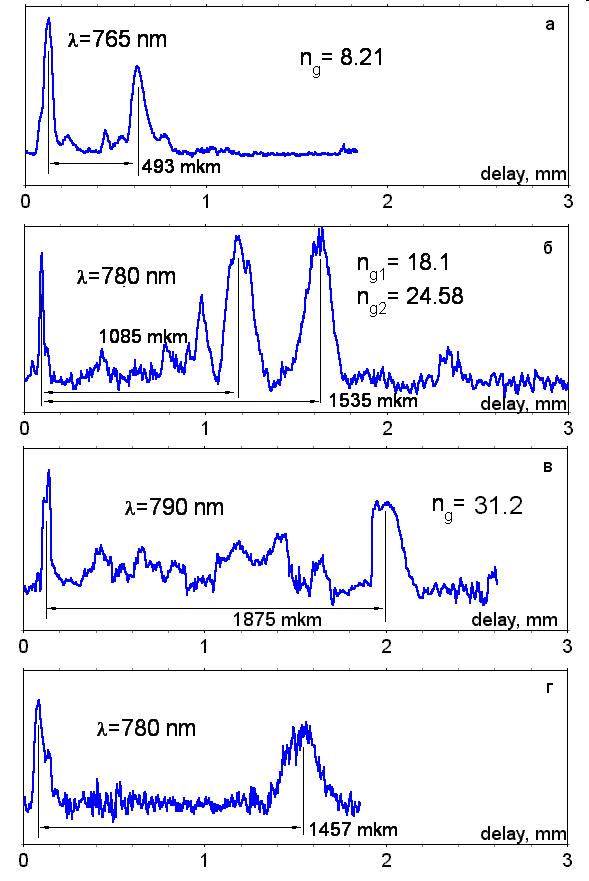}
 \caption{The delayed anti-mirror reflection for different wavelengths of the probe pulse. }
 \label{fig6}
 \end{center}
\end{figure}

\end{document}